\documentstyle[preprint,prb,aps,epsf,eqsecnum]{revtex}
\begin{document}
\tightenlines

\title{New Developments in the Eight Vertex Model}

\author{Klaus Fabricius
\footnote{e-mail Fabricius@theorie.physik.uni-wuppertal.de}}
\address{ Physics Department, University of Wuppertal, 
42097 Wuppertal, Germany}
\author{Barry~M.~McCoy
\footnote{e-mail mccoy@insti.physics.sunysb.edu}}               
\address{ Institute for Theoretical Physics, State University of New York,
 Stony Brook,  NY 11794-3840}
\date{\today}
\preprint{YITPSB-02-31}

\maketitle

\begin{abstract}

We demonstrate that the $Q$ matrix introduced in Baxter's 1972
solution of the eight vertex model has some eigenvectors  which are not 
eigenvectors of the spin reflection operator and 
conjecture a new functional equation for $Q(v)$
which both contains the Bethe equation that gives the eigenvalues of
the transfer matrix and computes the degeneracies of these eigenvalues.

\end{abstract}

\section{Introduction}

Thirty years ago Baxter \cite{bax0} computed the eigenvalues of
the transfer matrix of the eight vertex model in a paper of
unsurpassed brilliance and creativity. One of the key steps of
this method is the invention of an auxiliary matrix $Q(v)$ which
satisfies a functional equation with the transfer matrix $T(v)$. 

One year later Baxter computed the eigenvectors of the 
transfer matrix \cite{bax1}-\cite{bax3}
and in the course of that computation he again obtains the functional
equation between $T(v)$ and $Q(v)$ previously derived in ref.\cite{bax0}.
However, the definitions of $Q(v)$ used in ref.\cite{bax0} and in
ref.\cite{bax1}-\cite{bax3} are not the same and Baxter comments that
(page 15 of ref. \cite{bax1}) ``The above methods provide a different
(though obviously related) definition of $Q(v)$ to that of ref. \cite{bax0}
which may help us to understand $Q(v)$ a little better.''

We have been interested in extending our studies \cite{fm1}-\cite{fm4}
of the degeneracies
of the spectrum of the transfer matrix of the six vertex model at 
roots of unity to the eight vertex model and have seen in many
numerical examples that the exponentially degenerate multiplets of the
six vertex model also exist in the eight vertex model. 
In the course of the search for an explanation of these degeneracies
which would extend the $sl_2$ loop algebra symmetry of the six vertex
model to some analogous algebraic structure for the eight vertex model 
we have examined these two definitions of $Q(v)$ in detail. We have
discovered that while the two definitions are obviously related that the
two $Q(v){'\rm s}$ so defined are in fact different. 
First of all we find that there are cases where $Q_{72}(v)$ does not
exist. Furthermore in the case where $Q_{72}(v)$ exists
we have found that all the eigenvectors of $Q_{73}(v)$ of 
ref.\cite{bax1}-\cite{bax3}  and
ref. \cite{baxb} are eigenvectors of the spin reflection operator
whereas some of the
eigenvectors of the $Q_{72}(v)$ defined in ref. \cite{bax0} are not.
This lack of invariance under spin reversal of  $Q_{72}(v)$ 
does not affect the computation of the eigenvalues of
$T(v)$ of ref.\cite{bax0} but it does affect their degeneracy. 
Furthermore we have conjectured a functional equation for $Q_{72}(v)$ 
which  incorporates the ``Bethe equation'' whose roots
specify the eigenvalues of $T(v)$ and which also computes the degeneracy of
these eigenvalues by demonstrating that $Q_{72}(v)$ has zeroes specified by
the function introduced recently by Deguchi \cite{deg1}-\cite{deg2}.

In sec. 2 we review the formalism of ref.\cite{bax0}. In sec. 3 we present
our conjectured functional equation. We close in sec. 4 with a
discussion of the functional equation and its significance.

\section{Formalism of the Eight vertex model}

We use the notation of Baxter's 1972 paper \cite{bax0}.
The transfer matrix for the eight vertex model with $N$ columns and periodic
boundary conditions is
\begin{equation}
T_8(u)|_{\bf \mu,\nu}={\rm Tr} W_8(\mu_1,\nu_1)W_8(\mu_2,\nu_2)
\cdots W_8(\mu_N,\nu_N)
\label{t8}
\end{equation}
where in the conventions of  (6.2) of ref. \cite{bax0}
\begin{eqnarray}
W_8(1,1)|_{1,1}=W_8(-1,-1)|_{-1,-1}&=&~\rho
\Theta(2\eta)\Theta(v-\eta)
H(v+\eta)\nonumber\\
W_8(-1,-1)|_{1,1}=W_8(1,1)|_{-1,-1}&=&\rho\Theta(2\eta)
H(v-\eta)\Theta(v+\eta)\nonumber\\
W_8(-1,1)|_{1,-1}=W_8(1,-1)|_{-1,1}&=&\rho H(2\eta)
\Theta(v-\eta)\Theta(v+\eta)\nonumber\\
W_8(1,-1)|_{1,-1}=W_8(-1,1)|_{-1,1}&=&~\rho H(2\eta)
H(v-\eta)H(v+\eta).\nonumber\\
\label{bw}
\end{eqnarray}
The definition and useful properties of $H(v)$ and $\Theta(v)$ are
recalled in the appendix.

The computations of 
ref. \cite{bax0} are restricted to values of $\eta$ which
satisfy the ``root of unity condition'' (C15) of ref.\cite{bax0}.
We here further restrict our attention to the case which will connect
with our previous computations\cite{fm1}-\cite{fm4} in the six vertex model 
by setting in (C15) $m_2=0$ and thus obtaining
\begin{equation}
2L\eta=2m_1K.
\label{root}
\end{equation}

In the 1972 paper \cite{bax0} Baxter defines a matrix $Q_{72}(v)$ 
and states on page 200 of ref. \cite{bax0} that 
``... there are two elementary ways in which the matrix $T(v)$, and
hence the matrix $Q_{72}(v)$, can be broken up into diagonal blocks
or subspaces'' which are characterized by the quantum numbers $\nu '$
and $\nu''$ of (6.4) and (6.5) of ref. \cite{bax0}. 
The transfer matrix $T(v)$ certainly has this block diagonalization
property and if the transfer matrix $T(v)$ were non
degenerate the same property would follow for $Q_{72}(v)$  because $Q_{72}(v)$
commutes with the transfer matrix. But in the root of unity case (\ref{root})
the transfer matrix has degenerate eigenvalues and in these degenerate 
subspaces the eigenvectors of $Q_{72}(v)$ can fail to be eigenvectors
of the spin reflection operator.

In ref.\cite{bax0} the matrix
$Q_{72}(v)$ is explicitly defined by (C37)
\begin{equation}
Q_{72}(v)=Q_R(v)Q^{-1}_R(v_0)
\label{q72def}
\end{equation}
where $v_0$ is an arbitrary normalization point at which $Q_R(v)$ 
is nonsingular.
The matrix  $Q_R(v)$ in (\ref{q72def}) is defined as
\begin{equation}
[Q_R(v)]_{\alpha |\beta}={\rm Tr}S(\alpha_1, \beta_1)
S(\alpha_2, \beta_2)\cdots S(\alpha_N, \beta_N)
\end{equation}
where $\alpha_j$ and $\beta_j=\pm 1$ and $S(\alpha,\beta)$ is an
$L\times L$ matrix given as (C16),
\begin{equation}
S(\alpha,\beta)=\left( \begin{array}{cccccc}
z_0 & z_{-1}&0&0&\cdot&0\\
z_1 & 0  &z_{-2}&0&\cdot&0\\ 
0   &z_2 &0&z_{-3}&\cdot&0\\
\cdot&\cdot&\cdot&\cdot&\cdot&\cdot\\
0&0&0&\cdot&0&z_{1-L}\\
0&0&0&\cdot&z_{L-1}&z_L
\end{array}\right)
\end{equation}
with (C17)
\begin{equation}
z_m=q(\alpha, \beta, m|v)
\end{equation}
and (C19)
\begin{eqnarray}
q(+,\beta,m|v)&=&H(v+K+2m\eta)\tau_{\beta,m},\nonumber\\
q(-,\beta,m|v)&=&\Theta(v+K+2m\eta)\tau_{\beta,m}
\label{qtau}
\end{eqnarray} 
and we recall from (\ref{root}) that
\begin{equation}
\eta=m_1 K/L.
\end{equation}
The $\tau_{\beta,m}$ are generically arbitrary but we note that if
they are all set equal to unity then $Q_R(v)$ is so singular that
its rank becomes 1. On the other hand as long as the $\tau_{\beta,m}$
are chosen so that there is a $v_0$ such that $Q_R(v_0)$ is not singular
then $Q_{72}(v)$ is independent of $\tau_{\beta,m}$. 

In ref. \cite{bax0} it is stated that $Q_R(v)$ is nonsingular for
generic values of $v$ for any
$L$ for $N=1,2.$ We have extended these studies of the rank of
$Q_R(v)$ up to $N=9$ and $L=17$  and have found that 
the nonsingularity of $Q_R(v)$
breaks down if $L$ is odd and $m_1$ is even for sufficiently large $N.$
 The results for $2\leq L \leq 9$ are presented in table 1.

\vspace{.2in}

Table 1. Rank of the matrix $Q_R(v)$ for generic values of $v$ as a 
function of $L, m_1$ and $N.$
The ranks of the matrices which are singular are marked in bold face.
\vspace{.1in}

\begin{tabular}{|l|l|r|r|r|r|r|r|}\hline
$L$&$m_1$&rank $N=2$&rank $N=4$&rank $N=6$&rank N=7&rank $N=8$
&rank $N=9$\\ \hline
2&1&4&16&64&128&256&512\\ \hline
$3$&$1$&4&16&$64$&128&$256$&512\\
&$2$&{\bf3}&{\bf7}&{\bf18}&{\bf29}&{\bf47}&{\bf76}\\ \hline
$4$&1&4&16&64&128&256&512\\
   &3&4&16&64&128&256&512\\ \hline
5&1&4&16&64&128&256&512\\
 &2&4&{\bf13}&{\bf38}&{\bf57}&{\bf117}&{\bf193}\\
 &3&4&16&64&128&256&512\\
 &4&4&{\bf13}&{\bf38}&{\bf57}&{\bf117}&{\bf193}\\ \hline
6&1&4&16&64&128&256&512\\
 &5&4&16&64&128&256&512\\ \hline
7&1&4&16&64&128&256&512\\
 &2&4&16&{\bf57}&{\bf64}&{\bf187}&{\bf247}\\
 &3&4&16&64&128&256&512\\
 &4&4&16&{\bf57}&{\bf64}&{\bf187}&{\bf247}\\
 &5&4&16&64&128&256&512\\
 &6&4&16&{\bf57}&{\bf64}&{\bf187}&{\bf247}\\ \hline 
8&1&4&16&64&128&256&512\\
 &3&4&16&64&128&256&512\\
 &5&4&16&64&128&256&512\\
 &7&4&16&64&128&256&512\\ \hline
9&1&4&16&64&128&256&512\\
 &2&4&16&64&{\bf64}&{\bf248}&{\bf256}\\
 &4&4&16&64&{\bf64}&{\bf248}&{\bf256}\\
 &5&4&16&64&128&256&512\\
 &7&4&16&64&128&256&512\\
 &8&4&16&64&{\bf64}&{\bf248}&{\bf256}\\
\hline
\end{tabular}

\vspace{.2in}

There are several features of our study for $N\leq 9$ and $L\leq 17$ 
to be explicitly noted.

1) For $L$ even and for $L$ odd and $m_1$ odd the matrix $Q_R(v)$ is
  generically  nonsingular

2) For $L$ odd, $m_1$ even, and $N$ even $Q_R(v)$ is singular if
   $N\geq L-1.$ For $L=3$ and $N=2$ this contradicts the statement on
   nonsingularity on page 218 of ref. \cite{bax0}.

3) For $L$ odd, $m_1$ even and $N$ odd $Q_R(v)$ is singular for all
   $L.$ For $L\geq N$ the rank is $2^{N-1}$ which is one half the
   dimension of the matrix.

4) For even $N$ and all $L$ $Q_R(v)$ is singular at $v=0,K,iK$ and $K+iK';$
for odd $N$ and all $L$ $Q_R(v)$ is singular at $v=K$ and $K+iK'$ but
not at $0$ and $iK';$ for
even $N$ and $L>2$ $Q_R(v)$ is also singular at $v=\pm \eta$. 

The method of ref.\cite{bax0} assumes that $Q_R(v)$ is nonsingular and
   hence cannot literally hold in the cases where $L$ is odd and
   $m_1$ is even. However when $N$ is even we may use the symmetry of
   the transfer matrix eigenvalues $t(v;\eta)$
\begin{equation}
t(v+K;K-\eta)=(-1)^{\nu '}t(v;\eta)
\end{equation}
(where $\nu '=0,1$ is the quantum number (6.4) of ref. \cite{bax0}) 
to study the 
   singular case with $m_1$ even by transforming to the case
$m_1\rightarrow L-m_1$ where $Q_R(v)$ is nonsingular. 
In the rest of this paper
we restrict our attention to the cases where
$Q_{72}(v)$ exists.

From the definition the eigenvectors of $Q_{72}(v)$ may
be explicitly computed for small values of $N$ and $L.$ We have done
this for $L=2,3$, $m_1=1$ and $N=8$ and found that $Q_{72}(v)$ is
non degenerate and that in the subspaces where the eigenvalues of
$T(v)$ are degenerate there are eigenvectors of $Q_{72}(v)$ 
which are not eigenvectors of the spin reflection operator.

The failure of the eigenvectors of $Q_{72}(v)$ to all be eigenvectors
of the spin reflection operator means that the quantum number $\nu''$
of ref. \cite{bax0} can not in general be used for all eigenvectors of 
$Q_{72}(v).$ Therefore instead of the transformation properties (6.9) of  
ref.\cite{bax0} we have
\begin{eqnarray}
Q_{72}(v+2K)&=&(-1)^{\nu '}Q_{72}(v)\label{qper}\\
Q_{72}(v+2iK')&=&q^{-N}{\rm exp}(-iN\pi v/K)Q_{72}(v)
\label{qtrans}
\end{eqnarray}
where we note that (\ref{qtrans}) follows from the identity
\begin{equation}
S_R(\alpha,\beta|v+2iK')=q^{-1}e^{-i\pi v/K}MS_R(\alpha,\beta|v)M^{-1}
\end{equation}
with
\begin{equation}
M_{j,j'}=e^{-\pi i \eta j(j-1)/K}\delta_{j,j'}.
\end{equation}

From (\ref{qper}) and (\ref{qtrans}) one derives the most general 
form for the eigenvalues of
$Q_{72}(v)$ to be
\begin{equation}
Q_{72}(v)={\cal K}(q;v_k){\rm exp}(-i\nu\pi v/2K)\prod_{j=1}^{N}H(v-v_j)
\label{form1}
\end{equation}
where
\begin{eqnarray}
e^{i\pi (\nu'+\nu+N)}&=&1~~~{\rm so}~~\nu'+\nu+N={\rm even~integer}
\label{sum1}\\
e^{\pi i(-i \nu K'/K+N+\sum_{j=1}^Nv_j/K)}&=&1~~~
{\rm so}~~N+(-\nu iK'+\sum_{j=1}^Nv_j)/K={\rm even~integer}
\label{sum2}\nonumber\\
\end{eqnarray}
and ${\cal K}(q;v_k)$ is a normalization constant independent of $v.$
We choose by convention that the
$v_j$ lie in the fundamental region
\begin{equation}
0\leq Rev_j \leq 2K,~~~0\leq Im v_j \leq 2K.'
\end{equation}
The values of the even integers in the sum rules depend on the choice
of these conventions. 

From the imaginary part of (\ref{sum2}) we 
find an explicit formula for $\nu$
\begin{equation}
\nu=\sum_{j=1}^N Im v_j/K'
\label{eqnu1}
\end{equation}
and using this in (\ref{sum1}) we find
\begin{equation}
\nu=\sum_{j=1}^NIm v_j/K'=~~{\rm even ~~integer}-\nu'-N.
\label{eqnu2}
\end{equation}

From the real part of (\ref{sum2}) we obtain the sum rule
\begin{equation}
N+\sum_{j=1}^N Re v_j/K=~~{\rm even~~integer}
\label{sum3}
\end{equation}
We note that the difference between the form (\ref{form1}) and the
form (6.10) of ref. \cite{bax0} is that the imaginary period of the
fundamental region of
(6.10) is exactly half that of the fundamental region of (\ref{form1}).

For the cases $\eta=K/2,K/3$ ($L=2,3$) and $N=8$
the values of $v_j$ in (\ref{form1}) have been determined numerically and 
we find that not only are
the eigenvalues of $Q_{72}(v)$  
all of the form (\ref{form1}) but in fact can be written in the form  
\begin{eqnarray}
Q_{72}(v)={\cal K}(q;v_k){\rm exp}(-i\nu\pi v/2K)\prod_{j=1}^{n_B}
H(v-v^B_j)H(v-v^B_j-iK')\nonumber\\
\times\prod_{j=1}^{n_L}H(v-iw_j)H(v-iw_j-2K/L)\cdots H(v-iw_j-2(L-1)K/L)
\label{form2}
\end{eqnarray}
where
\begin{equation}
2n_B+Ln_L=N,
\end{equation}
the $w_l$ are real,
from (\ref{eqnu2}) $\nu$ is given by
\begin{equation}
\nu=n_B+(L\sum_{j=1}^{n_L}w_j+2\sum_{j=1}^{n_B}Im v^B_j)/K'
={\rm even~integer}-\nu'-N
\label{eqnu3}
\end{equation}
and from (\ref{sum3}) the $v^B_j$ satisfy the sum rule
\begin{equation}
N+n_L(L-1)+2\sum_{j=1}^{n_B}Re v_j^B/K={\rm even~integer}.
\label{sum4}
\end{equation}
We conjecture that the form (\ref{form2}) is correct for all even $N$
but we have explicitly seen that for odd $N$ it fails.


It is clear from (\ref{form2}) 
that there are two types of roots while
the form (6.10) of ref.\cite{bax0} incorporates only  one of these two types.
We remark that the zeros at
\begin{equation}
v=iw_j+2lK/L~~~l=0,\cdots,L-1
\label{new}
\end{equation}
are not the same as the zeroes of the form
\begin{equation}
v^B_j=v_j+2lK/L
\label{lstring}
\end{equation}
which have been called \cite{baxn} complete L-strings.
The zeroes of the form (\ref{new}) have not previously been seen.

Baxter shows in ref.\cite{bax0} that
the transfer matrix satisfies a functional equation (4.5) with (6.3)
\begin{equation}
T(v)Q_{72}(v)=[\rho h(v-\eta)]^N Q_{72}(v+2\eta)
+[\rho h(v+\eta)]^NQ_{72}(v-2\eta)
\label{funeqn}
\end{equation}
where
\begin{equation}
[T(v), Q_{72}(v')]=[T(v),T(v')]=[Q_{72}(v),Q_{72}(v')]
\label{comrel}
\end{equation}
and
\begin{equation}
h(v)=\Theta(0)\Theta(v)H(v)
\end{equation}
with 
\begin{eqnarray}
h(v+2K)&=&-h(v)\label{hper}\\
h(v+2iK')&=&q^{-2}e^{-2\pi i v/K}h(v).\label{htrans}
\end{eqnarray}
When our form (\ref{form2}) is put into
(\ref{funeqn}) we see that all the dependence on the $w_l$
cancels out and we are left with a ``Bethe's equation'' for the $v^B_j$
\begin{equation}
\left({h(v^B_l-\eta)\over h(v^B_l+\eta)}\right)^N
=e^{2\pi i \nu m_1/L}
\prod_{j=1\atop l\neq j}^{n_B}{h(v^B_l-v^B_j-2\eta)
\over h(v^B_l-v^B_j+2\eta)}
\label{beq8}
\end{equation}
where $\nu$ is given by (\ref{eqnu3}), the $v^B_j$ obey the sum 
rules (\ref{eqnu3}) and(\ref{sum4}) and
in the phase factor we have used the root of unity condition
(\ref{root}). When $n_L=0$ this reduces to equation (10.6.10) of Baxter's
book \cite{baxb} and this equation is commonly called ``Bethe's'' equation.
For this reason we call the $v^B_j$ Bethe roots.
We note that for $n_L\neq 0$ that (\ref{beq8}) has the same form as
Baxter's (10.6.10) but that both the $\nu$ given by (\ref{eqnu3}) 
which appears 
in the phase factor in (\ref{beq8}) and the sum rule in (\ref{eqnu3}) depend
on both $v^B_j$ and $w_j$ whereas the phase (10.6.7a) 
and the sum rule (10.6.7b)
of ref. \cite{baxb} depends only on $v_j^B.$

\section{Functional equation}

To complete the specification of the eigenvalues of $Q_{72}(v)$
we need to compute the $w_l.$ We wish to do this by producing a
functional equation satisfied by $Q_{72}(v).$ We take our inspiration
from

1) The polynomial $Y(v)$ (1.42) of our paper \cite{fm4}
\begin{equation}
Y(v)=\sum_{l=0}^{L-1}{\sinh^N{1\over 2}(v-(2l+1)i\gamma_0)
\over Q^B(v-2il\gamma_0)Q^B(v-2i(l+1)\gamma_0)}
\label{y1}
\end{equation}
where 
 \begin{equation}
Q^B(v)=\prod_{k=1}^n\sinh {1\over 2}(v-v^B_k)
\end{equation}
where the $v^B_k$ are the ``ordinary'' Bethe roots which do not include
any of the complete strings.
It is useful to rewrite this in the form
\begin{equation} 
{Y(v)Q^B(v)Q^B(v-2i\gamma_0)\over \sinh^N{1\over 2}(v-i\gamma_0)}
=\sum_{l=0}^{L-1}{\sinh^N {1\over 2}(v-(2l+1)i\gamma_0)\over 
\sinh^N {1\over 2}(v-i\gamma_0)}{Q(v)Q(v-2i\gamma_0)\over 
Q(v-2il\gamma_0)Q(v-2i(l+1)\gamma_0)}
\label{y2}
\end{equation}
where the  $Q(v)$ on the right is the full $Q(v)$ including all the
complete strings. The complete strings may be included here because
they cancel between the numerator and denominator.

2) The function of Deguchi in (5.8) of ref. \cite{deg1} and (31) of
  ref. \cite{deg2} which may be written as
\begin{equation}
G(v)=\sum_{l=0}^{L-1}(-1)^{rm_1l}e^{-4\eta cl}
{h^N(v-(2l+1)K/L)\over h^N(v-K/L)}
\prod_{k=1}^{n_B}{h(v-v^B_k)h(v-v^B_k-2K/L)\over
h(v-v^B_k-2lK/L)h(v-v^B_k-2(l+1)K/L)}
\end{equation}
where the $v^B_k$ are the ordinary roots and do not include complete strings.
This may easily be rewritten in terms of $Q_{72}(v)$ as given in 
(\ref{form2}) as
\begin{equation}
G(v)=\sum_{l=0}^{L-1}
{h^N(v-(2l+1)K/L)\over h^N(v-K/L)}
{Q_{72}(v)Q_{72}(v-2K/L)\over Q_{72}(v-2lK/L)Q_{72}(v-2(l+1)K/L)}
\label{g2}
\end{equation}

3) The functional equation for the transfer matrix $T_q$ of 
the 3 state chiral Potts model (2.21)
   of ref.\cite{amp}
\begin{eqnarray}
T_qT_{Rq}T_{R^2q}&=&K[({ab\over cd}\eta^2-1)^N({ab\over
dc}\eta^2\omega^2-1)^NT_q\nonumber\\
&+& ({ab\over cd}\eta^2\omega^2-1)^N({ab\over
cd}\eta^2\omega-1)^NT_{R^2q}\nonumber\\
&+&({ab\over cd}\eta^2-1)^N({ab\over
cd}\eta^2\omega-1)^N T_{R^4q}]
\label{cp}
\end{eqnarray}
where
$R$ is the automorphism (1.20) of ref. \cite{amp} (which is not to be
confused with the spin reflection operator).
This is better written by dividing by $T_qT_{R^2q}T_{R^4q}$
where the general form of (\ref{cp}) for general integer $L$ is
\begin{equation}
{T_{Rq}\over T_{R^{2(L-1)}q}}=K\sum_{l=0}^{L-1}
{f^N_l\over T_{R^{2l}q}T_{R^{2(l+1)}q}}
\label{cp2}
\end{equation}
with
\begin{equation}
f_l=({ab\over cd}\eta^2\omega^{-l}-1)({ab\over cd}\eta^2\omega^{1-l}-1)
\end{equation}
The form of the right hand side of (\ref{cp2}) is in the same form as the
right hand side of (\ref{y1}). If we divide by the term with $l=L-1$
we obtain a form 
with a right hand side comparable to(\ref{y2}) and
(\ref{g2})
\begin{equation}
{T_{Rq}T_q\over f^N_{L-1}}=K\sum_{l=0}^{L-1}{f^N_l\over 
f^N_{L-1}}{T_{R^{2(L-1)}q}
T_{q}\over T_{R^{2l}q}T_{R^{2(l+1)}q}}.
\label{t2}
\end{equation}

In order to make a conjecture we note that the right hand side of
(\ref{g2}) will agree with (\ref{t2}) if we replace the automorphism
$R^2$ by the shift $v\rightarrow v-2K/L$. The only other reasonable
replacement is to let in the left hand side of (\ref{t2}) the
automorphism $R$ be replaced by the automorphism $v\rightarrow v-iK'.$ 
We also note that any conjecture must be invariant under the transformation
$v\rightarrow v+2iK'.$
Thus we are led to the following

 \vspace{.2in}
{\bf CONJECTURE}
\vspace{.1in}

For $N$ even and either $L$ even or $L$ and $m_1$ odd  
\begin{eqnarray}
& &e^{-N\pi i v/2K}Q_{72}(v-iK')\nonumber\\
&=&A\sum_{l=0}^{L-1}h^N(v-(2l+1)K/L)
{Q_{72}(v)\over Q_{72}(v-2lK/L)Q_{72}(v-2(l+1)K/L)}
\label{con}
\end{eqnarray}
where $A$ is a normalizing  constant matrix independent of $v$ 
that commutes with $Q_{72}.$ What this matrix is depends
on the normalization value of $v_0$ in the definition (\ref{q72def}) 
of $Q_{72}$.  

\vspace {.1in}

There are several points to be noted about this conjecture.

1) The exponential factor in the left hand side is needed to maintain
invariance under the transformation $v \rightarrow v+2iK'$ as can be seen
by use of (\ref{htrans}) and (\ref{qtrans}). It is also needed to
insure that both sides are invariant under $v\rightarrow v+2K$ by use of
(\ref{hper}) and (\ref{qper}).  
With this factor 
both sides of the conjectured functional equation are  
quasi periodic functions with the same fundamental region.

2) If we multiply out the denominators this
conjecture may be  rewritten as
\begin{eqnarray}
& &e^{-N\pi i v/2K}
Q_{72}(v-iK')\prod_{l=1}^{L-1}Q_{72}(v-2lK/L)\nonumber\\
&=&A\{\sum_{l=0}^{L-2}h^N(v-(2l+1)K/L)Q_{72}(v)\cdots
Q_{72}(v-2(l-1)K/L)Q_{72}(v-2(l+2)K/L)\nonumber\\
& &\cdots Q_{72}(v-2(L-1)K/L)\nonumber\\
&+&(-1)^{\nu'}h^N(v-(2L-1)K/L)\prod_{l=1}^{L-2}Q_{72}(v-2lK/L)\}
\end{eqnarray}
In this form the conjecture has been  proven for $L=2$ and all even $N$
and has been numerically verified for $L=3,$ $m_1=1$ and $N=8.$

3) If we use (\ref{form2}) in (\ref{con}) we obtain
\begin{eqnarray}
& &(-1)^{n_B+n_L}q^{(\nu/2 - n_{B})}\exp(-{i\pi \over
K}\{(\nu-n_{B}+N/2)v +\sum_{j=1}^{n_B}v^{B}_{j}\})\nonumber\\
&\times&
\prod_{j=1}^{n_L}\prod_{l=0}^{L-1}H(v-iw_j-2lK/L)H(v-i(w_j+K')-2lK/L)
\nonumber \\
&=&A{\cal K}(q;v^B_k)^{-2} 
\sum_{l=0}^{L-1}{e^{-i\nu\pi (2l+1)/L}h^N(v-(2l+1)K/L)
\over \prod_{j=1}^{n_B}H(v-2lK/L-v_j^{B})H(v-2lK/L-v_j^{B}-iK')}\nonumber\\
&\times&{1\over \prod_{j=1}^{n_B}
H(v-2(l+1)K/L-v_j^{B})H(v-2(l+1)K/L-v_j^{B}-iK')}
\label{wv}
\end{eqnarray}
Here we note that the left hand side depends only on $w_l$ and the right hand side
only on $v^{B}_l$.

4) The apparent poles in (\ref{wv}) when
$v=v^B_k+2lK/L,v^B_k+2lK/L+iK'$ cancel
   because the $v^B_k$ are specified  by the Bethe equation
   (\ref{beq8}). This is exactly what we saw in the expression
   (\ref{y1}) for the
   polynomial $Y(v)$ in ref. \cite{fm4}. 
Equivalently we may say that the Bethe's equation for
$v^B_k$ is already included in (\ref{con}).

5) The left hand side of (\ref{wv}) is symmetric under 
exchange of $w_l$ and $w_l+K'$ and all  theta functions 
$H$ appear in pairs $H(u)$, $H(u-iK')$
which can be combined to $h(u)$.
Therefore for any given set of Bethe roots  
$v^{B}_k$ the $n_L$ equations in(\ref{wv}) 
have $2^{n_L}$ independent solutions for the $w_l$ 
which thus determines the
dimensionality of the multiplet of degenerate 
eigenstates of the transfer matrix.

6) In the XXZ limit the right hand side of (\ref{wv}) reduces to the
polynomial $Y(v)$ (1.42) of ref. \cite{fm4} once the possibility of
Bethe roots at infinity is taken into account
. The zeroes of $Y(v)$ have
been identified with the evaluation parameters of the loop $sl_2$
symmetry algebra of the XXZ model in ref. \cite{fm4}.

\section{Discussion}

The conjectured functional equation (\ref{con}) provides an elegant
computation of the  $2^{n_L}$ degeneracy of the eigenvalues of the
transfer matrix of the eight vertex model previously   
found in our numerical computations.
In the six vertex model limit these powers of 2 are explained by
showing that only spin one half representations occur in the decomposition
into irreducible representations of the loop $sl_2$ symmetry algebra. 
The functional equation (\ref{con}) obtains this result 
without any reference to a symmetry algebra. Such a derivation of the
eigenvalue multiplicities is superior to the derivation from the loop
algebra because the loop algebra symmetry of the six vertex model has
only been analytically demonstrated \cite{fm1} for the 
case $S^z\equiv 0~~({\rm mod}~L).$

In the above we have written $Q_{72}(v)$ in terms of the $Q_R(v)$
by use of the first equation in (C37) of ref.\cite{bax0}. 
In order to  use the corresponding formulas with $Q_L(v)$ the
right hand sides of the  two
equations in (C26) need to be interchanged. This modification of (C26) 
is consistent with the subsequent equations in appendix C.

At present the functional equation (\ref{con}) is a conjecture which
requires proof.
Because of the similarity of 
(\ref{con}) to the functional equation of the chiral Potts model
it is natural to investigate whether the techniques used in the chiral
Potts model can be extended to the eight vertex model. The proof of the
chiral Potts functional equation is given in the work of Bazhanov and
Stroganov \cite{bs} and Baxter, Bazhanov and Perk \cite{bbp}. However,
it is instantly apparent from these papers that there exists the
possibility that there may be even more  matrices beyond $Q_{72}(v)$
and $Q_{73}(v)$ which satisfy
Baxter's functional equation (\ref{funeqn}).
In particular for the six vertex model is shown on page 805 on ref. \cite{bs}
that there is a five parameter family of $Q(v)$ matrices (denoted by
$\cal T$) which satisfy
Baxter's functional equation and at the very least it seems plausible
that both $Q_{72}(v)$ and $Q_{73}(v)$ should be embedded in such a 
larger family.
This reflects the fact that all methods of solution of the 8 vertex model 
include steps which, while they are sufficient to obtain the transfer
matrix eigenvalues, are quite probably not necessary. 
As an example we note that the condition
\begin{equation}
[Q(v),Q(v')]=0
\label{qcomm}
\end{equation}
required by Baxter \cite{bax0}-\cite{bax3} for both $Q_{72}(v)$ 
and $Q_{73}(v)$ can be replaced by the weaker condition
\begin{equation}
[Q(v),Q(v\pm2\eta)]=0
\label{qetacom}
\end{equation}
which is necessary to reduce the functional equation for the matrices
$T(v)$ and $Q(v)$
(\ref{funeqn}) 
to an equation for eigenvalues.
The work of ref. \cite{bs} shows that matrix $\cal T$ 
satisfies (\ref{qetacom}) but that in general (\ref{qcomm}) does not hold.
It is surprising
that after 30 years these questions have not been resolved.

\appendix

\section{Theta functions}

The definition of Jacobi Theta functions of nome $q$ is
\begin{eqnarray}
H(v)&=&2\sum_{n=1}^\infty(-1)^{n-1}q^{(n-{1\over 2})^2}\sin[(2n-1)\pi
v/(2K)]\\
\Theta(v)&=&1+2\sum_{n=1}^{\infty}(-1)^nq^{n^2}\cos(nv\pi/K)\nonumber\\
&=&-iq^{1/4}e^{\pi i v/(2K)}H(v+iK')
\label{thetadf}
\end{eqnarray}
where $K$  and $K'$ are the standard elliptic integrals of the first kind
and 
\begin{equation}
q=e^{-\pi K'/ K}.
\end{equation}
These theta functions satisfy the quasi periodicity relations (15.2.3) of ref. \cite{baxb}
\begin{eqnarray}
H(v+2K)&=&-H(v)\label{hper1}\\
H(v+2iK')&=&-q^{-1}e^{-\pi i v/K}H(v)\label{Hper}
\end{eqnarray}
and
\begin{eqnarray}
\Theta(v+2K)&=&\Theta(v)\\
\Theta(v+2iK')&=&-q^{-1}e^{-\pi i v/ K}\Theta(v).
\end{eqnarray}
>From (\ref{thetadf}) we see that $\Theta(v)$ and $H(v)$ are not
independent but satisfy
(15.2.4) of ref.\cite{baxb}
\begin{eqnarray}
\Theta (v+iK')=iq^{-1/4}e^{-{\pi i v\over 2K}}H(v)\nonumber \\
H(v+iK')=iq^{-1/4}e^{-{\pi i v\over 2K}}\Theta(v).
\label{threl}
\end{eqnarray}

\vspace{.2in}

\centerline{\bf ACKNOWLEDGMENTS}

\vspace{.2in}

We are pleased to thank Prof. R.J. Baxter, Prof. V.V. Bazhanov 
and Prof. V. Jones for fruitful
discussions. One of us (BMM) is pleased to thank the Miller Institute
and the mathematics department of the University of California at
Berkeley for their hospitality. This work is partially supported by
NSF grant DMR 0078058.


\begin{thebibliography}{99}



\bibitem{bax0} R.J. Baxter, Partition function of the eight vertex
model, Ann. Phys. 70 (1972) 193.


\bibitem{bax1}R.J. Baxter,
Eight--vertex model in lattice statistics and one--dimensional
anisotropic Heisenberg chain I: Some fundamental eigenvectors,
Ann. Phys. 76 (1973) 1.

\bibitem{bax2}R.J. Baxter,
Eight--vertex model in lattice statistics and one--dimensional
anisotropic Heisenberg chain II: Equivalence to a generalized
Ice-type lattice model, Ann. Phys. 76 (1973) 25.


\bibitem{bax3}R.J. Baxter, Eight--vertex model in
lattice statistics and one--dimensional
anisotropic Heisenberg chain III: Eigenvectors of the  transfer matrix
and the  Hamiltonian,
Ann. Phys. 76 (1973)48.

\bibitem{fm1} T. Deguchi, K. Fabricius and B.M. McCoy, The $sl_2$ loop
algebra symmetry of the six-vertex model at roots of
unity. J. Stat. Phys. 102 (2001) 701.

\bibitem{fm2} K. Fabricius and B.M. McCoy, Bethe's equation is
incomplete for the XXZ model at roots of unity, J. Stat. phys. 103
(2001) 647.

\bibitem{fm3} K. Fabricius and B.M. McCoy, Completing Bethe's
equations at roots of unity, J. Stat. Mech. 104 (2001) 573.

\bibitem{fm4} K. Fabricius and B.M. McCoy, Evaluation parameters and
Bethe roots for the six vertex model at roots of unity, in Progress
in Mathematical Physics vol 19, ed. M. Kashiwara and T. Miwa
(Birkhauser, Boston 2002) 119.


\bibitem{baxb}R. J. Baxter, {\it Exactly solved models}, Academic Press.
London (1982).




\bibitem{deg1} T. Deguchi, Construction of some missing
eigenvectors of the XYZ spin chain at discrete coupling constants
and the exponentially large spectral degeneracy of the transfer matrix,
J. Phys. A 35 (2002) 879.

\bibitem{deg2} T. Deguchi, The 8V CSOS model and the $sl_2$
loop algebra symmetry of the six-vertex model at roots of unity,
Int. J. Mod. Phys. B 16 (2002) 1899.

\bibitem{baxn}R. J. Baxter, On the completeness of the Bethe
Ansatz for the six and eight-vertex models
J. Stat. Phys. 108 (2002) 1.

\bibitem{amp} G. Albertini, B.M. McCoy and J.H.H. Perk, Eigenvalue
spectrum of the superintegrable chiral Potts model, Adv. Stud. in Pure
Math. 19 (1989) 1. 



\bibitem{bs} V.V. Bazhanov and Yu. G. Stroganov, Chiral Potts model as
a descendant of the six-vertex model, J. Stat. Phys. 59 (1990) 799. 

\bibitem{bbp}R.J. Baxter, V.V. Bazhanov and J.H.H.Perk, Functional
relations for transfer matrices of the chiral Potts model,
Int. J. Mod. Phys. B4 (1990) 803.  


\end{thebibliography}
\end{document}